\documentclass[sigconf]{acmart}

\usepackage{multirow}
\usepackage{float}
\usepackage{placeins}
\usepackage{subfig}
\usepackage{enumitem}
\usepackage{algorithmic}
\usepackage[ruled, linesnumbered]{algorithm2e}
\usepackage{pgfplots}
\usepgfplotslibrary{groupplots}
\pgfplotsset{compat=1.18}

\AtBeginDocument{%
  }

\setcopyright{none}
\renewcommand\footnotetextcopyrightpermission[1]{}

\title{CaIRec: Calibrated Modality Imputation for Incomplete Multimodal Recommendation}

\newcommand{\fullname}{\textbf{\textit{Ca}}librated \textbf{\textit{I}}mputation for Incomplete Multimodal \textit{\textbf{Rec}}ommendation}
\newcommand{\shortname}{\textsc{\textbf{\textit{CaIRec}}}}
\newlength{\savedcolumnwidth}

\author{Ruiyu Liu}
\authornote{This work was completed while Ruiyu Liu was a research intern at the National University of Singapore.}
\affiliation{%
  \institution{Southern University of Science and Technology}
  \city{Shenzhen}
  \country{China}
  }
\email{12310410@mail.sustech.edu.cn}

\author{Xiaohao Liu}
\affiliation{%
  \institution{National University of Singapore}
  \city{Singapore}
  \country{Singapore}
  }
\email{xiaohao.liu@u.nus.edu}

\author{Miaomiao Cai}
\authornote{Corresponding author.}
\affiliation{%
  \institution{National University of Singapore}
  \city{Singapore}
  \country{Singapore}
  }
\email{miaomiao.cai@nus.edu.sg}

\author{Yunshan Ma}
\affiliation{%
  \institution{Singapore Management University}
  \city{Singapore}
  \country{Singapore}
  }
\email{ysma@smu.edu.sg}

\author{See-Kiong Ng}
\affiliation{%
  \institution{National University of Singapore}
  \city{Singapore}
  \country{Singapore}
  }
\email{seekiong@nus.edu.sg}

\begin{document}

\begin{abstract}

Real-world multimodal recommender systems frequently suffer from incomplete modality observations, where items lack images, text, or other important content features. Such incompleteness weakens item representations and can substantially degrade recommendation performance. Existing modality imputation methods estimate missing representations from the available item content, but two key challenges remain. First, these methods mainly optimize the recovered representation itself, without explicitly considering how it should relate to the other modalities of the same item. The completed modalities may therefore form inconsistent cross-modal relations, causing \textbf{Cross-modal Structural Distortion}. Second, even when the completed modalities are structurally coherent, the recovered information may still be ineffective for personalized ranking. Recovered representations receive limited ranking-oriented guidance, while modality missingness also disrupts the item neighborhoods required for preference propagation, resulting in a \textbf{Preference Adaptation Gap}. 

To address these challenges, we propose ~\fullname~(\shortname), a two-stage framework for incomplete multimodal recommendation. In the first stage, \textbf{\textit{Structural Imputation Calibration}} (SIC) estimates missing-modality representations from shared information inferred from the available modalities and calibrates their cross-modal organization through structural regularization and correspondence supervision from genuinely observed modality pairs. In the second stage, \textbf{\textit{Preference-oriented Representation Calibration}} (PRC) performs recommendation-specific adaptation at both the representation and relation levels. It constructs pseudo-missing instances to align recovered representations with their observed counterparts shaped by ranking supervision in the recommendation space. It further builds completion-aware item graphs by integrating completed content relations with collaborative evidence. Extensive experiments on three real-world datasets under different modality-missing settings demonstrate the effectiveness and robustness of ~\shortname.

\end{abstract}

\maketitle
\setlength{\textfloatsep}{10pt plus 1pt minus 2pt}
\setlength{\floatsep}{8pt plus 1pt minus 2pt}
\setlength{\intextsep}{10pt plus 1pt minus 2pt}
\setlength{\dbltextfloatsep}{10pt plus 1pt minus 2pt}
\setlength{\dblfloatsep}{8pt plus 1pt minus 2pt}
\setlength{\savedcolumnwidth}{\columnwidth}

\section{Introduction}

Multimodal recommendation improves item understanding by using different types of content information, such as images and text~\cite{he2016vbpr,chen2017attentive,kang2017visually,wei2019mmgcn,tao2020mgat,wei2023lightgt}. Different modalities describe an item from different views and provide useful information beyond user interactions~\cite{zhang2021lattice,liu2022disentangled,wang2023dualgnn,liu2023semantic,zhou2023bootstrap,li2024attribute}. Such information is especially important when interactions are sparse~\cite{wei2019mmgcn,jiang2024diffmm}. However, most existing methods assume that all items have complete modality information~\cite{wang2018lrmm,malitesta2024drop,kim2025disentangling}. This assumption often does not hold in real-world applications, where some items may lack images, text descriptions, or other content features~\cite{lin2023contrastive,ganhor2024sibrar}. Previous studies have shown that such modality incompleteness can substantially reduce the performance of multimodal recommendation models~\cite{wang2018lrmm,ganhor2024sibrar,malitesta2024dealing}. Therefore, incomplete multimodal recommendation aims to learn reliable item representations and user preferences from partial modality observations for accurate recommendation.

Existing studies mainly address incomplete multimodal recommendation through robust representation learning and modality imputation. Robust methods, such as MILK~\cite{bai2024multimodality} and $I^3$MRec~\cite{chen2025i3mrec}, reduce the sensitivity of recommendation models to different missing patterns by extracting stable preference signals or modality-invariant information from the available modalities. However, their content evidence remains limited to the observed modalities, since they do not explicitly estimate the unavailable modality representations. In contrast, modality imputation directly estimates unavailable features or representations~\cite{lin2023contrastive,malitesta2026trainingfree,li2026robust}. For example, MoDiCF~\cite{li2025generating} generates missing modality features through a diffusion process, while DGMRec~\cite{kim2025disentangling} integrates shared semantics, modality-specific information, and user preferences for representation recovery. By estimating unavailable modality representations, they improve incomplete item representations and recommendation performance. 

\begin{figure}[t]
    \centering
    \includegraphics[width=\linewidth]{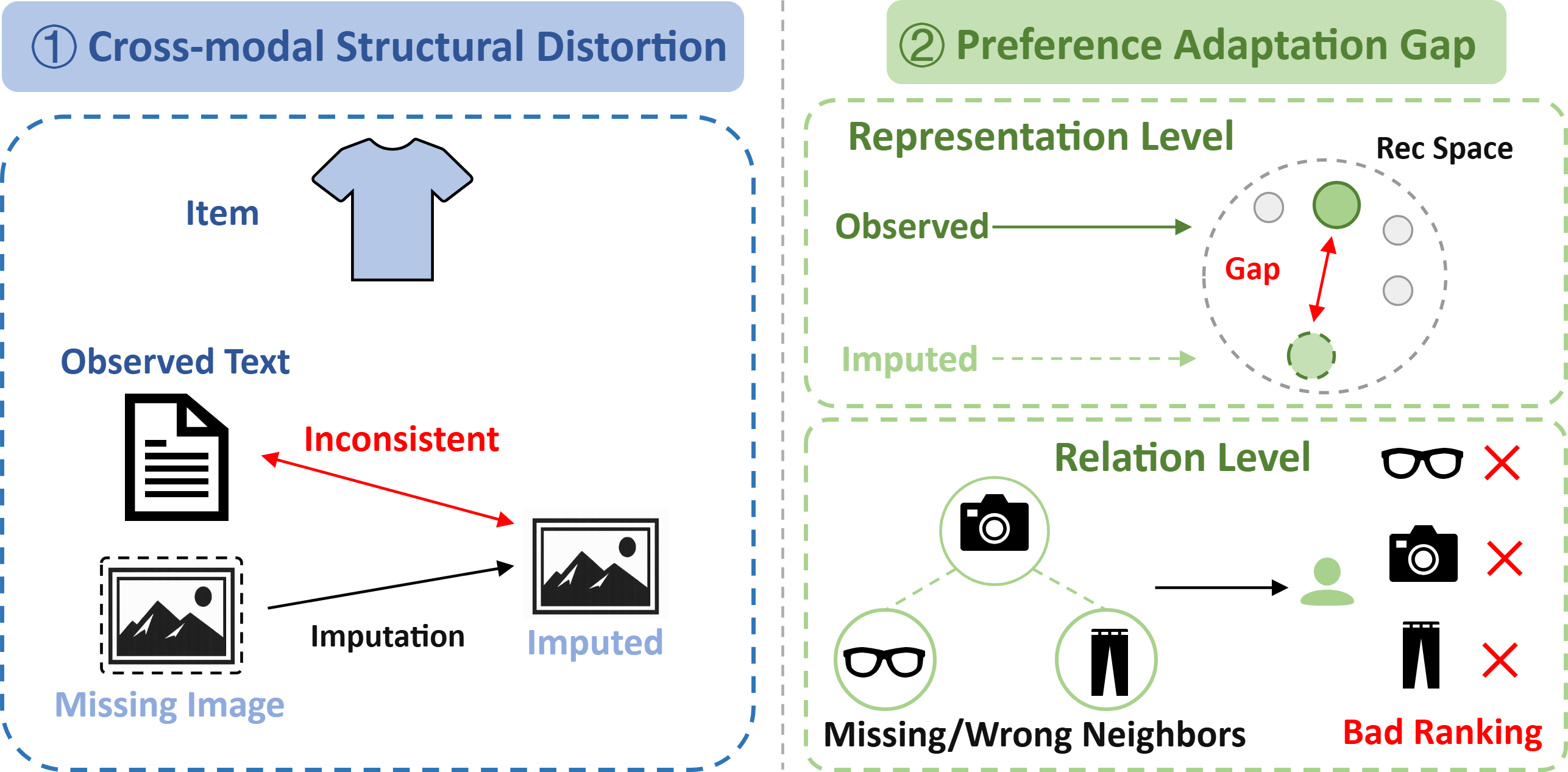}
    \caption{The two challenges in incomplete multimodal recommendation, \textit{Cross-modal Structural Distortion} and the \textit{Preference Adaptation Gap}.}
    \label{fig:introduction_motivation}
\end{figure}

Despite these advances, existing imputation methods mainly recover representations from the available content of each item, while paying less attention to their cross-modal organization and effectiveness for personalized ranking. As illustrated in Figure~\ref{fig:introduction_motivation}, this limitation gives rise to two key challenges. First, \textbf{Cross-modal Structural Distortion} arises because a missing modality is commonly estimated from the observed modalities of the same item. Although such item-wise recovery can capture predictable information, the recovered representation is not explicitly constrained by its relations with the other modalities. It may therefore deviate from the cross-modal relation patterns exhibited by genuinely observed modality pairs, resulting in a distorted organization of the completed modalities. Second, the \textbf{Preference Adaptation Gap} arises because structurally reliable completion does not directly ensure effectiveness for recommendation. Existing imputation methods mainly optimize content recovery~\cite{li2025generating,kim2025disentangling,li2026robust}, while personalized ranking requires additional adaptation under interaction-based supervision. At the representation level, recovered modalities follow a different generation pathway from genuine content and may deviate from their ranking-supervised observed counterparts in the recommendation space. At the relation level, modality missingness removes content-derived item connections, while item-wise recovery does not necessarily reconstruct the neighborhoods required for preference propagation. These representation- and relation-level mismatches form the Preference Adaptation Gap. Together, these challenges require structurally calibrated modality completion followed by recommendation-oriented adaptation.

To address these challenges, we propose ~\fullname~(\shortname), a two-stage framework for incomplete multimodal recommendation. Building on recent advances in multimodal completion~\cite{li2026robust,liu2026calibrated}, ~\shortname~ first employs \textbf{\textit{Structural Imputation Calibration (SIC)}} to establish a structurally calibrated completion foundation and then applies \textbf{\textit{Preference-oriented Representation Calibration (PRC)}} to adapt completed representations and item relations to personalized ranking. Specifically, SIC estimates missing-modality representations from available item content and calibrates their cross-modal organization. Its structural calibration regularizes within-item modality relations, while correspondence calibration anchors the completion space to cross-modal correspondence learned from genuinely observed modality pairs. More importantly, PRC performs recommendation-specific adaptation at both representation and relation levels. At the representation level, it constructs pseudo-missing instances to align recovered representations with ranking-supervised observed counterparts in the recommendation space. At the relation level, it combines completed content relations with collaborative evidence to supplement item neighborhoods disrupted by modality missingness. Through these two stages, ~\shortname~ first improves the structural reliability of modality completion and then transforms completed information into recommendation-effective signals. Extensive experiments on three datasets under different modality-missing settings demonstrate the effectiveness and robustness of ~\shortname.
The main contributions of this work are summarized as follows:

\begin{itemize}[leftmargin=0.5cm, itemindent=0cm]

    \item We identify two successive challenges in incomplete multimodal recommendation. \textit{Cross-modal Structural Distortion} concerns the cross-modal organization of recovered modalities, while the \textit{Preference Adaptation Gap} concerns their suitability for personalized ranking.

    \item We propose ~\fullname~(\shortname), a two-stage framework connecting modality completion with recommendation. SIC calibrates the within-item cross-modal structure, while PRC adapts recovered representations and item relations to preference learning.

    \item Extensive experiments on three real-world datasets under diverse modality-missing settings demonstrate the effectiveness and robustness of ~\shortname.

\end{itemize}

\section{Preliminaries}

Let $\mathcal U=\{1,\ldots,N_u\}$ and $\mathcal I=\{1,\ldots,N_i\}$ denote the user and item sets, respectively. The training interactions are represented by $\mathbf R\in\{0,1\}^{N_u\times N_i}$, where $R_{ui}=1$ indicates an observed interaction and $R_{ui}=0$ denotes an unobserved interaction rather than an explicit negative preference~\cite{hu2008implicit,rendle2009bpr}.
Multimodal recommendation further incorporates heterogeneous item content, such as visual and textual features, which may be incomplete in practice~\cite{he2016vbpr,wang2018lrmm}. Let $\mathcal M=\{1,\ldots,N_m\}$ denote the modality set. For each item $i$, $\mathcal O_i\subseteq\mathcal M$ and $\mathcal Q_i=\mathcal M\setminus\mathcal O_i$ denote its observed and missing modality sets, respectively. We assume $\mathcal O_i\neq\varnothing$. For each $m\in\mathcal O_i$, item $i$ has a pre-trained feature $\mathbf x_i^m\in\mathbb R^{d_m}$. When $m\in\mathcal Q_i$, the corresponding feature is unavailable and is not replaced by a zero-filled observation.

Given $\mathbf R$ and the partially observed modality features $\mathcal X^{\mathrm{obs}}=\{\mathbf x_i^m\mid i\in\mathcal I,m\in\mathcal O_i\}$, incomplete multimodal recommendation aims to learn user and item representations $\mathbf h_u,\mathbf h_i\in\mathbb R^d$. The preference score is computed as~\cite{he2020lightgcn}: 
\begin{equation}
    \hat y_{ui}=\mathbf h_u^\top\mathbf h_i.
    \label{eq:prediction}
\end{equation}
Candidate items are ranked based on $\hat y_{ui}$. The key challenge is to learn effective user and item representations from partial modality observations for accurate personalized ranking.

\section{Method}

\begin{figure*}[t]
    \centering
    \includegraphics[width=\linewidth]{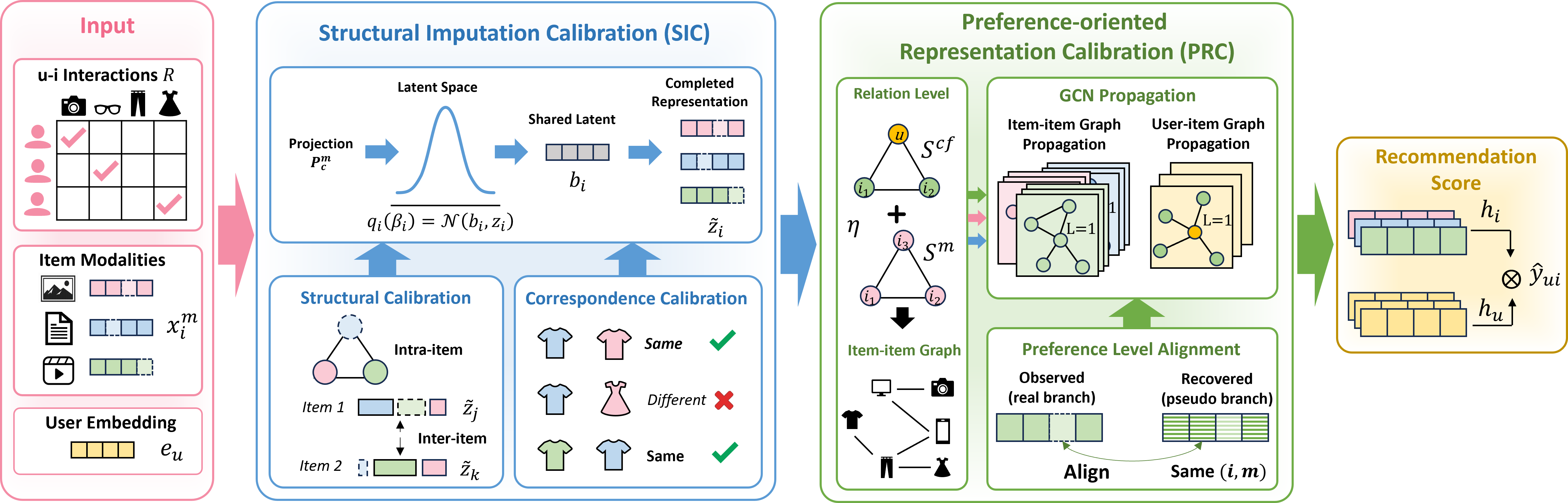}
    \caption{Overview of the proposed ~\shortname{} framework. SIC estimates missing-modality representations and calibrates their within-item cross-modal organization. PRC then adapts recovered representations in the recommendation space and supplements item relations for preference propagation and personalized ranking.}
    \label{fig:framework}
\end{figure*}

As illustrated in Figure~\ref{fig:framework}, we propose ~\fullname~(\shortname), a two-stage framework comprising \textbf{\textit{Structural Imputation Calibration (SIC)}} and \textbf{\textit{Preference-oriented Representation Calibration (PRC)}}. Building on calibrated representation imputation, SIC first estimates missing-modality representations from shared information inferred from the available item content. It then calibrates the cross-modal organization of the completed modalities through structural regularization and correspondence supervision from genuinely observed modality pairs. The observed and recovered modality representations are retained as explicit completed item content for subsequent recommendation. PRC further performs recommendation-specific adaptation at both the representation and relation levels. It constructs pseudo-missing instances to reduce the discrepancy between recovered and observed representation pathways in the recommendation space, and builds completion-aware item graphs by integrating completed content relations with collaborative evidence. Through these two stages, ~\shortname{} first establishes a structurally calibrated completion space and then adapts the completed representations and item relations to personalized ranking.

\subsection{Structural Imputation Calibration}

To address \textbf{\textit{Cross-modal Structural Distortion}}, SIC follows an impute-then-calibrate strategy. Existing imputation methods focus on recovering individual missing-modality representations from available content, while paying less attention to their relations with the other modalities of the same item. SIC first estimates missing-modality representations through a shared-latent imputation backbone and then introduces structural and correspondence calibration to reduce the resulting cross-modal relation inconsistency. 

\subsubsection{Shared-latent imputation backbone}
Since modality features differ in dimensions and distributions, we map them into a common completion space. For each modality $m$, we employ a modality-specific projection $P_c^m:\mathbb R^{d_m}\rightarrow\mathbb R^{d_c}$. For observed modality $m\in\mathcal O_i$, its projected representation is defined as:
\begin{equation}
    \mathbf z_i^m = \frac{P_c^m(\mathbf x_i^m)} {\|P_c^m(\mathbf x_i^m)\|_2+\epsilon}, \qquad m\in\mathcal O_i,
    \label{eq:sic_projection}
\end{equation}
where $d_c$ denotes completion dimension and $\epsilon>0$ ensures stability.

Building on representation imputation and probabilistic latent-variable modeling~\cite{tipping1999probabilistic,bach2005probabilistic,klami2015group,ghojogh2021factor,liu2026calibrated}, we instantiate an item-specific shared latent model for modality imputation. Each item $i$ is associated with a shared latent variable $\boldsymbol{\beta}_i\in\mathbb R^{d_\beta}$, with prior $p(\boldsymbol{\beta}_i)=\mathcal N(\mathbf 0,\mathbf I)$. Conditioned on $\boldsymbol{\beta}_i$, modality $m$ is modeled as: 
\begin{equation}
    p(\mathbf z_i^m\mid\boldsymbol\beta_i) = \mathcal N \left( \mathbf W^m\boldsymbol\beta_i+\boldsymbol\mu^m, (\sigma_m)^2\mathbf I \right),
    \label{eq:sic_generation}
\end{equation}
where $\mathbf W^m$ and $\boldsymbol\mu^m$ map the shared latent information into modality $m$, and $\sigma_m$ models the residual uncertainty. This model combines item-level shared information with modality-specific generation mappings, allowing missing representations to be conditionally estimated in their corresponding modality spaces.

To aggregate the information available for item $i$, SIC estimates the posterior of $\boldsymbol\beta_i$ conditioned on the observed projected representations $\{\mathbf z_i^m\mid m\in\mathcal O_i\}$ and the generative parameters $\Theta_g=\{\mathbf W^m,\boldsymbol\mu^m,\sigma_m\}_{m\in\mathcal M}$. The posterior combines evidence from all observed modalities and has the following closed form~\cite{bach2005probabilistic,klami2015group,ghojogh2021factor}: 
\begin{equation}
    q_i(\boldsymbol\beta_i) = p\left( \boldsymbol\beta_i \mid \{\mathbf z_i^m\}_{m\in\mathcal O_i}, \Theta_g \right) = \mathcal N \left( \mathbf b_i, \boldsymbol\Sigma_i \right),
    \label{eq:sic_posterior}
\end{equation}
where $\mathbf b_i$ and $\boldsymbol\Sigma_i$ are the posterior mean and covariance. Because the posterior depends only on the observed modality set $\mathcal O_i$, the same inference process applies to different missing patterns. We use $\mathbf b_i$ for missing-modality estimation and provide the closed-form expressions in Appendix~\ref{app:sic_inference}. 

To learn the shared latent factors and modality-specific generation patterns from the available observations, we optimize the following representation objective~\cite{ghojogh2021factor}:
\begin{small}
\begin{equation}
    \mathcal L_{\mathrm{rep}}=-\frac{1}{|\mathcal B|}\sum_{i\in\mathcal B}\left[\mathbb E_{q_i(\boldsymbol\beta_i)}\left[\sum_{m\in\mathcal O_i}\log p(\mathbf z_i^m\mid\boldsymbol\beta_i)\right]-D_{\mathrm{KL}}\left(q_i(\boldsymbol\beta_i)\|p(\boldsymbol\beta_i)\right)\right],
    \label{eq:sic_representation_objective}
\end{equation}
\end{small}
where $\mathcal B$ is the current item batch. The reconstruction term encourages the latent variable to explain the observed modality representations, while the KL term regularizes its posterior distribution. Only modalities in $\mathcal O_i$ participate in the objective, so missing modalities are never treated as zero-valued observations. 

After estimating the item-specific latent information, we map the posterior mean $\mathbf b_i$ into each missing modality space through the corresponding generative parameters~\cite{bach2005probabilistic,klami2015group,ghojogh2021factor}. For each $m\in\mathcal Q_i$, the recovered representation is: 
\begin{equation}
    \hat{\mathbf z}_i^m=\frac{\mathbf W^m\mathbf b_i+\boldsymbol\mu^m}{\|\mathbf W^m\mathbf b_i+\boldsymbol\mu^m\|_2+\epsilon}.
    \label{eq:sic_imputation}
\end{equation}
This representation is a conditional estimate rather than an exact reconstruction of the unavailable content. The posterior mean $\mathbf b_i$ summarizes information predictable from the observed modalities, while $\mathbf W^m$ and $\boldsymbol\mu^m$ transfer it into the target modality space through generation patterns learned across items. The imputation backbone therefore estimates only the target information supported by the available observations. Unlike related calibration-based approaches that use imputation to recalibrate the observed modalities, SIC treats the recovered representation as missing-modality content and calibrates its cross-modal organization to mitigate Cross-modal Structural Distortion. For unified notation, we define $\tilde{\mathbf z}_i^m=\mathbf z_i^m$ for $m\in\mathcal O_i$ and $\tilde{\mathbf z}_i^m=\hat{\mathbf z}_i^m$ for $m\in\mathcal Q_i$. 

\subsubsection{Cross-modal structural calibration}

The shared-latent backbone estimates each missing representation from the available item content. However, individual recovery does not explicitly constrain how the recovered representation should relate to the other modalities of the same item. It may therefore deviate from the cross-modal relation patterns supported by valid modality observations, resulting in \textit{Cross-modal Structural Distortion}. To address this issue, we introduce structural and correspondence calibration to regularize the cross-modal organization of the completed modalities. 

For each item $i$, we stack its completed modality representations as $\tilde{\mathbf Z}_i= [\tilde{\mathbf z}_i^1,\ldots,\tilde{\mathbf z}_i^{N_m}] \in\mathbb R^{d_c\times N_m}$ and construct the modality-level Gram matrix $\mathbf G_i=\tilde{\mathbf Z}_i^\top\tilde{\mathbf Z}_i$. Since all representations are normalized, each entry of $\mathbf G_i$ measures the similarity between two modalities of the same item. We decompose the Gram matrix as $\mathbf G_i=\mathbf U_i\boldsymbol\Lambda_i\mathbf U_i^\top$, where $\lambda_{i,1}\geq\cdots\geq\lambda_{i,N_m}$ are the eigenvalues and $\mathbf u_i=[\mathbf U_i]_{:,1}$ is the principal eigenvector. The leading eigenvalue measures the concentration of the within-item modality relations, while the principal eigenvector characterizes the relative contribution of each modality to the principal relation pattern. 

Based on this spectral characterization~\cite{liu2026principled}, we regularize the completed representations from two perspectives. The leading-eigenvalue term reduces ambiguity in the within-item relation pattern, while the principal-direction term discourages identical structural configurations across items when the modality set admits nontrivial directional variation: 
\begin{equation}
    \mathcal L_{\mathrm{str}}=-\frac{1}{|\mathcal B|}\sum_{i\in\mathcal B}\left[\log\frac{\exp(\lambda_{i,1}/\tau)}{\sum_{r=1}^{N_m}\exp(\lambda_{i,r}/\tau)}+\log\frac{\exp(1/\tau)}{\sum_{j\in\mathcal B}\exp((\mathbf u_i^\top\mathbf u_j)^2/\tau)}\right],
    \label{eq:sic_structural_loss}
\end{equation}
where $\tau>0$ is the temperature hyperparameter. The first term increases the relative prominence of the leading eigenvalue and encourages a more concentrated cross-modal organization. The second term regularizes structural variation across items, while the squared inner product removes the sign ambiguity of eigenvectors.

In the bimodal case, the principal-direction term introduces no additional continuous gradient within a fixed cross-modal similarity sign pattern. The leading-eigenvalue term therefore provides the main structural optimization signal in our setting. We provide a detailed analysis in Appendix~\ref{app:bimodal_structure}. 

Structural calibration regularizes the overall relation pattern of each completed item, but it does not explicitly determine whether representations from different modalities correspond to the same item~\cite{chen2023vast,liu2026principled,cicchetti2025triangle}. To provide more direct cross-modal correspondence supervision, we construct a set $\mathcal P_{\mathcal B}$ of observed cross-modal pairs $p=((i,m),(j,n))$, where $m\neq n$. A pair is labeled positive when $i=j$ and negative otherwise. Positive pairs therefore contain two observed modalities of the same item, while negative pairs preserve the modality combination but replace one representation with that of another item. Based on these pairs, we introduce the following correspondence calibration objective: 
\begin{equation}
    \mathcal L_{\mathrm{corr}}=-\frac{1}{|\mathcal P_{\mathcal B}|}\sum_{p\in\mathcal P_{\mathcal B}}\left[y_p\log \hat y_p+(1-y_p)\log(1-\hat y_p)\right],
    \label{eq:sic_correspondence_loss}
\end{equation}
where $y_p$ is the correspondence label and $\hat y_p= \sigma(g^{m,n}([\mathbf z_i^m\Vert\mathbf z_j^n]))$ is the predicted probability. Only genuinely observed modality pairs are used, which avoids treating uncertain imputed representations as supervision targets. Because the modality projections and generative model share a completion space, this objective provides correspondence supervision from genuinely observed modality pairs and anchors the space to cross-modal correspondence.  

Finally, we combine the imputation and structural calibration objectives as:
\begin{equation}
    \mathcal L_{\mathrm{SIC}} = \mathcal L_{\mathrm{rep}} + \lambda_{\mathrm{str}}\mathcal L_{\mathrm{str}} + \lambda_{\mathrm{corr}}\mathcal L_{\mathrm{corr}},
    \label{eq:sic_objective}
\end{equation}
where $\lambda_{\mathrm{str}}$ and $\lambda_{\mathrm{corr}}$ control the two calibration objectives. The representation loss learns the shared-latent imputation backbone. The structural loss regularizes the spectral organization of the completed modalities, while the correspondence loss anchors the completion space to observed cross-modal correspondence. 

\subsection{Preference-oriented Representation Calibration}

Although SIC produces structurally calibrated representations in the completion space, such representations are not necessarily well suited to personalized ranking. This limitation arises at both the representation and relation levels. At the representation level, recovered and directly observed representations follow different processing pathways and may therefore exhibit a pathway-specific discrepancy in the recommendation space. At the relation level, modality missingness removes content-derived item connections, while item-wise completion does not necessarily reconstruct the neighborhoods needed for preference propagation. To address these issues, we introduce \textbf{Preference-oriented Representation Calibration (PRC)} as the recommendation-specific stage of ~\shortname{}. PRC adapts the recovery pathway through pseudo-missing instances and constructs completion-aware item graphs by integrating collaborative evidence with completed content relations. 

\subsubsection{Representation-level adaptation}

Although recovered modality representations participate in the final ranking objective, ranking supervision alone provides no paired reference for correcting the pathway-specific discrepancy introduced by modality recovery. For the same item and modality, the recovered representation is inferred from the available modalities, whereas its observed counterpart is directly extracted from the original content. These different processing pathways may cause the recovered representation to deviate from its observed counterpart in the recommendation space, leaving it insufficiently adapted to personalized ranking. To provide direct adaptation guidance, we construct pseudo-missing instances from observed modalities. Specifically, we mask an observed modality, recover it from the remaining modalities, and pair the resulting pseudo-recovered representation with its observed counterpart in the recommendation space. Since each pair shares the same item and modality, their discrepancy mainly reflects the deviation introduced by the recovery pathway.

For modality $m$, we map observed and recovered representations through separate branches. We employ a recommendation projection $P_r^m:\mathbb R^{d_c}\rightarrow\mathbb R^d$ for observed representations and an adapter $A^m:\mathbb R^{d_c}\rightarrow\mathbb R^d$ for recovered representations. The recommendation-space representation is defined as:
\begin{equation}
    \mathbf s_i^m =
    \begin{cases}
        P_r^m(\mathbf z_i^m), & m\in\mathcal O_i,\\
        A^m(\hat{\mathbf z}_i^m), & m\in\mathcal Q_i.
    \end{cases}
    \label{eq:prc_mapping}
\end{equation}
The observed branch is derived from observed modality content and is directly shaped by ranking supervision, making it a suitable reference in the recommendation space. The recovered branch follows the imputation pathway and may exhibit a pathway-specific shift. We therefore use a separate adapter instead of forcing both branches to share the same transformation. 

A genuinely missing modality has no observed representation that can serve as an alignment reference. We therefore construct pseudo-missing instances from items with at least two observed modalities. The eligible item--modality pairs are: 
\begin{equation}
    \mathcal E_{\mathrm{pm}}=\left\{(i,m)\mid i\in\mathcal I,\, m\in\mathcal O_i,\, |\mathcal O_i|\geq2 \right\}
\end{equation}
For each $(i,m)\in\mathcal E_{\mathrm{pm}}$, modality $m$ is temporarily masked and recovered from $\mathcal O_i\setminus\{m\}$ using the same SIC imputation process. We denote the pseudo-recovered representation by $\hat{\mathbf z}_i^{m,(-m)}$. Because the masked modality remains available as observed content during training, $\mathbf z_i^m$ provides a paired observed reference. The correspondence and pseudo-missing objectives require a nonempty subset of items with at least two observed modalities. When such paired observations are unavailable, these objectives cannot be directly instantiated, and the framework reduces to the shared-latent imputation and relation-level recommendation components. 

We align the pseudo-recovered and observed representations after mapping them into the recommendation space: 
\begin{equation}
    \mathcal L_{\mathrm{align}}=\frac{1}{|\mathcal E_{\mathrm{pm}}|} \sum_{(i,m)\in\mathcal E_{\mathrm{pm}}} \left[1-\cos\left(A^m(\hat{\mathbf z}_i^{m,(-m)}),\operatorname{sg}\left(P_r^m(\mathbf z_i^m)\right)\right)\right],
    \label{eq:prc_alignment}
\end{equation}
where $\operatorname{sg}(\cdot)$ denotes stop-gradient. Because each pair shares the same item and modality identity, the objective mainly captures the discrepancy introduced by the recovery pathway. Minimizing $\mathcal L_{\mathrm{align}}$ adapts the pseudo-recovered branch toward the ranking-supervised observed branch. Stop-gradient prevents the alignment loss from changing the observed reference, which remains optimized through personalized ranking. Pseudo-recovered and genuinely missing representations share the same SIC imputation process and modality-specific adapter. The learned pathway adaptation can therefore be applied to genuinely missing modalities.  

\subsubsection{Relation-level reconstruction}

Representation-level adaptation calibrates recovered modalities for individual items, but recommendation also depends on relations among items. When modality $m$ is missing, the corresponding content-derived connections cannot be obtained from the original observations. This leaves the modality-specific item neighborhood incomplete and weakens the associated preference propagation paths. Moreover, recovering an individual representation does not by itself ensure that the resulting content neighborhood is consistent with collaborative preference patterns. Motivated by modality-aware item-graph modeling~\cite{zhang2021lattice} and graph-based feature propagation under missing modalities~\cite{malitesta2024dealing,malitesta2026trainingfree}, we construct a completion-aware item graph integrating completed content and collaborative similarities. 

We model item relations from both interaction patterns and completed modality content. Let $d_i=\sum_{u\in\mathcal U}R_{ui}$ denote the interaction degree of item $i$. For two distinct items $i$ and $j$, the collaborative and modality-specific content similarities are: 
\begin{equation}
    S_{ij}^{\mathrm{cf}} = \frac{\sum_{u\in\mathcal U}R_{ui}R_{uj}} {\sqrt{d_i d_j}+\epsilon}, \qquad S_{ij}^{m}=\max\left(0,\cos\left(\tilde{\mathbf z}_i^m, \tilde{\mathbf z}_j^m\right)\right).
    \label{eq:item_similarity}
\end{equation}
Collaborative similarity measures normalized overlap between user interaction histories, while content similarity measures item proximity under modality $m$. Because $\tilde{\mathbf z}_i^m$ may be either observed or recovered, all items can participate in modality-specific relation construction. We set $S_{ii}^{\mathrm{cf}}=S_{ii}^m=0$ to exclude self-connections. 

The two relation sources provide complementary evidence. Collaborative similarity introduces interaction-supported preference proximity but may be unreliable for items with sparse interactions. Completed content similarity supplements modality-specific relations but may contain uncertainty when recovered representations are involved. We therefore fuse them as: 
\begin{equation}
    \mathbf S^{m,\mathrm{fuse}}=(1-\eta)\mathbf S^{\mathrm{cf}}+\eta\mathbf S^m,
    \label{eq:fused_similarity}
\end{equation}
where $\eta\in[0,1]$ balances the two relation sources. For each item, we select the Top-$k$ entries in $\mathbf S^{m,\mathrm{fuse}}$ as its neighbors and construct $\mathbf G_{\mathrm{item}}^m$. The resulting graph combines completed content relations with interaction-supported evidence, supplementing item connections disrupted by modality missingness and providing additional paths for preference propagation.

\subsubsection{Preference propagation and personalized ranking}

For each modality $m$, we augment the user--item interaction graph with the completion-aware item graph $\mathbf G_{\mathrm{item}}^m$. We then perform modality-specific preference propagation following LightGCN~\cite{he2020lightgcn} and existing multimodal graph recommendation methods~\cite{wei2019mmgcn,zhang2021lattice,guo2024lgmrec}. The propagated representations are aggregated across layers and modalities to obtain user and item embeddings $\mathbf h_u$ and $\mathbf h_i$. Their preference score is computed according to Equation~\eqref{eq:prediction}. The propagation and aggregation equations are provided in Appendix~\ref{app:graph_propagation}. 

We optimize the final representations using Bayesian Personalized Ranking~(BPR)~\cite{rendle2009bpr}. Let $\mathcal D_{\mathrm{BPR}} = \{(u,i,j)\mid R_{ui}=1,R_{uj}=0\}$ denote the training triplets, where user $u$ has interacted with item $i$ but not with item $j$. The recommendation loss is: 
\begin{equation}
    \mathcal L_{\mathrm{rec}}=-\frac{1}{|\mathcal D_{\mathrm{BPR}}|}\sum_{(u,i,j)\in\mathcal D_{\mathrm{BPR}}}\log\sigma\left(\hat y_{ui}-\hat y_{uj}\right).
    \label{eq:prc_rec}
\end{equation}

The overall objective of PRC combines personalized ranking, recovery-pathway alignment, and parameter regularization: 
\begin{equation}
    \mathcal L_{\mathrm{PRC}}=\mathcal L_{\mathrm{rec}}+\lambda_{\mathrm{align}}\mathcal L_{\mathrm{align}}+\lambda_{\mathrm{reg}} \lVert\Theta_{\mathrm{PRC}}\rVert_2^2,
    \label{eq:prc_objective}
\end{equation}
where $\lambda_{\mathrm{align}}$ and $\lambda_{\mathrm{reg}}$ control alignment and regularization terms.

We optimize SIC and PRC sequentially to separate structurally calibrated modality completion from recommendation-oriented adaptation. SIC is first trained with $\mathcal L_{\mathrm{SIC}}$ to learn the imputation backbone and organize the completed modalities in the completion space. We then fix the SIC parameters and optimize PRC with $\mathcal L_{\mathrm{PRC}}$. This stage adapts the recommendation-space mappings and supplements item relations without changing the learned imputation process. Accordingly, SIC addresses \textit{Cross-modal Structural Distortion} by calibrating cross-modal organization of conditionally imputed representations, while PRC addresses the \textit{Preference Adaptation Gap} through representation-level pathway adaptation and relation-level neighborhood reconstruction. Further theoretical and empirical analyses of efficiency and scalability are in Appendix~\ref{app:model_analysis}.

\begin{table}[t]
  \centering
  \caption{Statistics of three datasets.}
  \label{tab:dataset_statistics}
  \begin{tabular}{c|cccc}
    \toprule
    \textbf{Dataset} & \#\textbf{Users} & \#\textbf{Items} & \#\textbf{Interactions} & \textbf{Sparsity} \\
    \midrule
    \textbf{Clothing}       & 39,387  & 23,033 & 278,677 & 99.97\% \\
    \textbf{Sports}         & 35,598  & 18,357 & 296,337 & 99.95\% \\
    \textbf{Beauty}         & 21,752  & 11,161 & 165,325 & 99.93\% \\
    \bottomrule
  \end{tabular}%
\end{table}

\FloatBarrier
% \raggedbottom
\section{Experiments}

\subsection{Experimental Settings}

\begin{table*}[t]
  \centering
  \caption{Performance comparison under a 50\% modality-missing rate. Results are averaged over five runs. The best and second-best results are highlighted in bold and underlined, respectively. $^*$ indicates a statistically significant improvement over the strongest baseline under a paired two-sided $t$-test with $p<0.005$.}
  \label{tab:main_50_missing}
  \resizebox{\textwidth}{!}{
  \begin{tabular}{l|cccc|cccc|cccc}
    \toprule
    \multirow{2}{*}{\textbf{Methods}} 
    & \multicolumn{4}{c|}{\textbf{Clothing}}
    & \multicolumn{4}{c|}{\textbf{Sports}}
    & \multicolumn{4}{c}{\textbf{Beauty}} \\
    \cmidrule(lr){2-5} \cmidrule(lr){6-9} \cmidrule(lr){10-13}
    & R@10 & N@10 & R@20 & N@20
    & R@10 & N@10 & R@20 & N@20
    & R@10 & N@10 & R@20 & N@20 \\
    \midrule
    LightGCN  & 0.0301 & 0.0163 & 0.0453 & 0.0201 & 0.0526 & 0.0290 & 0.0814 & 0.0364 & 0.0440 & 0.0216 & 0.0696 & 0.0280 \\
    SimGCL    & 0.0402 & 0.0219 & 0.0604 & 0.0270 & 0.0639 & 0.0351 & 0.0974 & 0.0439 & 0.0477 & 0.0245 & 0.0764 & 0.0320 \\
    \midrule
    VBPR      & 0.0332 & 0.0184 & 0.0491 & 0.0224 & 0.0550 & 0.0298 & 0.0840 & 0.0373 & 0.0478 & 0.0233 & 0.0776 & 0.0308 \\
    BM3       & 0.0405 & 0.0220 & 0.0602 & 0.0270 & 0.0634 & 0.0346 & 0.0959 & 0.0430 & 0.0475 & 0.0237 & 0.0747 & 0.0305 \\
    PGL       & 0.0413 & 0.0222 & 0.0615 & 0.0273 & 0.0571 & 0.0306 & 0.0895 & 0.0390 & 0.0422 & 0.0211 & 0.0675 & 0.0275 \\
    MIG-GT    & \underline{0.0524} & \underline{0.0286} & \underline{0.0776} & \underline{0.0350} & \underline{0.0664} & \underline{0.0366} & \underline{0.1011} & \underline{0.0455} & \underline{0.0520} & \underline{0.0254} & 0.0804 & \underline{0.0325} \\
    \midrule
    I$^3$-MRec & 0.0416 & 0.0224 & 0.0647 & 0.0283 & 0.0620 & 0.0337 & 0.0951 & 0.0423 & 0.0474 & 0.0235 & 0.0768 & 0.0309 \\
    MILK      & 0.0284 & 0.0150 & 0.0433 & 0.0188 & 0.0298 & 0.0158 & 0.0475 & 0.0203 & 0.0340 & 0.0170 & 0.0543 & 0.0221 \\
    MoDiCF    & 0.0474 & 0.0266 & 0.0712 & 0.0329 & 0.0613 & 0.0346 & 0.0935 & 0.0432 & 0.0477 & 0.0237 & 0.0759 & 0.0308 \\
    DGMRec    & 0.0505 & 0.0273 & 0.0762 & 0.0339 & 0.0662 & 0.0361 & 0.1007 & 0.0450 & 0.0508 & 0.0247 & \underline{0.0810} & 0.0323 \\
    HEAT      & 0.0504 & 0.0273 & 0.0767 & 0.0340 & 0.0655 & 0.0358 & 0.0982 & 0.0443 & 0.0500 & 0.0248 & 0.0788 & 0.0320 \\
    \midrule
    \textbf{\shortname{} (Ours)} & \textbf{0.0537*} & \textbf{0.0291*} & \textbf{0.0814*} & \textbf{0.0361*} & \textbf{0.0709*} & \textbf{0.0383*} & \textbf{0.1058*} & \textbf{0.0474*} & \textbf{0.0529*} & \textbf{0.0261*} & \textbf{0.0842*} & \textbf{0.0339*} \\
    \textbf{Improv.}    & 2.48\% & 1.75\% & 4.90\% & 3.14\% & 6.78\% & 4.64\% & 4.65\% & 4.18\% & 1.73\% & 2.76\% & 3.95\% & 4.31\% \\
    \bottomrule
  \end{tabular}
  }
\end{table*}

\subsubsection{Datasets}

We evaluate ~\shortname{} on three Amazon recommendation datasets, namely \textbf{Clothing}, \textbf{Sports}, and \textbf{Beauty}~\cite{mcauley2015image}. Each dataset contains user--item ratings together with visual and textual item features. Following prior multimodal recommendation studies~\cite{zhou2023bootstrap}, we treat observed ratings as implicit feedback and adopt the same preprocessing and evaluation settings. We use the released 4,096-dimensional CNN visual features~\cite{mcauley2015image,zhou2023bootstrap} and 384-dimensional Sentence-Transformer textual features~\cite{zhou2023bootstrap,reimers2019sentencebert}. The dataset statistics are reported in Table~\ref{tab:dataset_statistics}.

\subsubsection{Modality-Missing Settings}
Following previous studies on incomplete multimodal recommendation~\cite{chen2025i3mrec,kim2025disentangling,li2026robust}, we simulate modality incompleteness by randomly selecting $50\%$ of the items as incomplete. For each selected item, either its visual or textual modality is uniformly sampled and removed, while the other modality remains observed. The remaining $50\%$ of the items retain both modalities. The resulting item-level missing mask is fixed across training, validation, and testing, so that each item has the same modality availability throughout an experiment. The removed features are unavailable to all methods, and all compared methods use the same missing mask for a fair comparison.

\subsubsection{Baselines, Evaluation Protocol and Implementation Details}
We compare ~\shortname{} with representative methods from three categories. (1) \textit{Collaborative Filtering} methods rely primarily on user--item interactions for preference modeling, including LightGCN~\cite{he2020lightgcn} and SimGCL~\cite{yu2022simgcl}. (2)\textit{General Multimodal Recommendation} methods incorporate visual and textual item content to enhance preference modeling, including VBPR~\cite{he2016vbpr}, BM3~\cite{zhou2023bootstrap}, PGL~\cite{yu2025pgl}, and MIG-GT~\cite{hu2025miggt}. (3) \textit{Incomplete Multimodal Recommendation} methods explicitly address modality missingness. Robustness-oriented methods learn stable representations from the available modalities, including $I^3$MRec~\cite{chen2025i3mrec} and MILK~\cite{bai2024multimodality}. Completion-based methods estimate unavailable modality features or representations, including MoDiCF~\cite{li2025generating}, DGMRec~\cite{kim2025disentangling}, and HEAT~\cite{malitesta2026trainingfree}. Detailed descriptions of all compared methods are provided in Appendix~\ref{app:baselines}.  

We follow the data splitting and full-ranking evaluation protocol of prior multimodal recommendation studies~\cite{zhou2023bootstrap}. Performance is measured by Recall@$K$ (R@$K$) and NDCG@$K$ (N@$K$) for $K\in\{10,20\}$, with all non-interacted items treated as candidates. Each experiment is repeated five times with different random seeds while keeping the data split and modality-missing mask fixed, and average results are reported. All methods use the same split, mask, and hyperparameter search protocol, with significance assessed by a paired two-sided $t$-test. Detailed implementation settings and search ranges are provided in Appendix~\ref{app:hyperparameter_settings}. 

% We follow the data splitting and ranking evaluation protocol used in prior multimodal recommendation studies~\cite{zhou2023bootstrap}. We evaluate recommendation performance using Recall@$K$(R@$K$) and NDCG@$K$(N@$K$), where $K=\{10,20\}$. For each user, all non-interacted items are treated as candidate items during evaluation. We repeat each experiment five times with different random seeds while keeping the data split and modality-missing mask fixed, and report the average results. All methods use the same fixed split and mask to ensure a fair comparison. Statistical significance is evaluated using a paired two-sided $t$-test. 
% All methods use the same data splits and item-level modality-missing patterns. Their hyperparameters are selected according to validation performance under the same search protocol. Detailed implementation settings, hyperparameter search ranges, and selected values are in Appendix~\ref{app:hyperparameter_settings}. 

% \vspace{-0.5cm}
\subsection{Overall Performance}

Table~\ref{tab:main_50_missing} reports the overall results on the three datasets. Across the evaluated datasets and metrics, ~\shortname{} consistently achieves the strongest performance. The results show that ~\shortname{} can effectively use partially observed multimodal content and adapt the completed representations and item relations to personalized ranking. We summarize the main observations as follows: 

\begin{itemize}[leftmargin=*,topsep=2pt,itemsep=2pt,parsep=0pt]

  \item \textbf{Completion-based methods generally outperform robustness-oriented methods.} $I^3$MRec and MILK improve robustness by extracting stable or modality-invariant information from the available modalities, but they do not explicitly estimate the unavailable modality representations. Their content evidence therefore remains limited to the observed modalities. In comparison, completion-based methods such as DGMRec and HEAT generally achieve stronger results than the robustness-oriented baselines. This comparison suggests that robustness to missing patterns alone may be insufficient, while explicit completion can provide additional content evidence for recommendation. 

  \item \textbf{MIG-GT remains a strong baseline under modality missingness.} Although MIG-GT is not specifically designed for incomplete modalities, it models visual, textual, and collaborative signals through separate branches and captures global dependencies with a Transformer. When one content modality is missing, its remaining content and collaborative branches can still provide useful recommendation signals. This design helps MIG-GT outperform several missing-aware methods. However, MIG-GT does not explicitly estimate missing-modality representations or perform recommendation-specific adaptation for the recovered pathway, and it therefore remains behind ~\shortname. 

  \item \textbf{\shortname{} achieves the strongest overall performance.} Across all dataset--metric combinations, ~\shortname{} yields an average relative improvement of 3.77\% over the strongest competing baseline. It also consistently surpasses leading completion-based methods, including DGMRec and HEAT. These results suggest that modality completion alone may not fully address incomplete multimodal recommendation. Individually recovered representations may still form distorted cross-modal relations and remain insufficiently adapted to personalized ranking. By combining Structural Imputation Calibration with Preference-oriented Representation Calibration, ~\shortname{} more effectively transforms completed modality information into recommendation-oriented representations and item relations. 

\end{itemize}

\subsection{Ablation Study}

\begin{table}[t]
  \centering
  \caption{Ablation study on three datasets, with relative drops
  from the full model shown in parentheses.}
  \label{tab:ablation_amazon}
  \newcommand{\dropcell}[2]{%
    \begin{tabular}[c]{@{}c@{}}
      #1\\[-0.2ex]
      {\scriptsize($-$#2\%)}
    \end{tabular}
  }
  \resizebox{\columnwidth}{!}{%
  \begin{tabular}{l|cccccc}
    \toprule
    \multirow{2}{*}{\textbf{Variant}}
      & \multicolumn{2}{c}{\textbf{Clothing}}
      & \multicolumn{2}{c}{\textbf{Sports}}
      & \multicolumn{2}{c}{\textbf{Beauty}} \\
    \cmidrule(lr){2-3}
    \cmidrule(lr){4-5}
    \cmidrule(lr){6-7}
      & R@20 & N@20
      & R@20 & N@20
      & R@20 & N@20 \\
    \midrule
    \textbf{\shortname{}}
      & \textbf{0.0814} & \textbf{0.0361}
      & \textbf{0.1058} & \textbf{0.0474}
      & \textbf{0.0842} & \textbf{0.0339} \\
    \midrule
    \textbf{w/o StrCal}
      & \dropcell{0.0772}{5.16}
      & \dropcell{0.0345}{4.43}
      & \dropcell{0.1049}{0.85}
      & \dropcell{0.0470}{0.84}
      & \dropcell{0.0819}{2.73}
      & \dropcell{0.0331}{2.36} \\
    \textbf{w/o CorrCal}
      & \dropcell{0.0803}{1.35}
      & \dropcell{0.0358}{0.83}
      & \dropcell{0.1050}{0.76}
      & \dropcell{0.0470}{0.84}
      & \dropcell{0.0833}{1.07}
      & \dropcell{0.0334}{1.47} \\
    \textbf{w/o SIC}
      & \dropcell{0.0756}{7.13}
      & \dropcell{0.0339}{6.09}
      & \dropcell{0.1032}{2.46}
      & \dropcell{0.0468}{1.27}
      & \dropcell{0.0797}{5.34}
      & \dropcell{0.0323}{4.72} \\
    \midrule
    \textbf{w/o PMA}
      & \dropcell{0.0792}{2.70}
      & \dropcell{0.0354}{1.94}
      & \dropcell{0.1044}{1.32}
      & \dropcell{0.0467}{1.48}
      & \dropcell{0.0828}{1.66}
      & \dropcell{0.0331}{2.36} \\
    \textbf{w/o ItemGraph}
      & \dropcell{0.0752}{7.62}
      & \dropcell{0.0334}{7.48}
      & \dropcell{0.1017}{3.88}
      & \dropcell{0.0450}{5.06}
      & \dropcell{0.0802}{4.75}
      & \dropcell{0.0322}{5.01} \\
    \textbf{w/o PRC}
      & \dropcell{0.0729}{10.44}
      & \dropcell{0.0326}{9.70}
      & \dropcell{0.1002}{5.29}
      & \dropcell{0.0444}{6.33}
      & \dropcell{0.0788}{6.41}
      & \dropcell{0.0314}{7.37} \\
    \bottomrule
  \end{tabular}%
  }
\end{table}

To analyze the contribution of the components in SIC and PRC, we construct six variants of ~\shortname{}. For SIC, \textit{w/o StrCal} and \textit{w/o CorrCal} remove the structural and correspondence calibration objectives, respectively, while \textit{w/o SIC} removes the entire SIC stage and represents each missing modality with a zero vector. For PRC, \textit{w/o ItemGraph} removes the completion-aware item graph, \textit{w/o PMA} removes pseudo-missing alignment, and \textit{w/o PRC} directly uses SIC outputs for recommendation without representation-level pathway adaptation or relation-level neighborhood reconstruction. Table~\ref{tab:ablation_amazon} reports the results under a $50\%$ modality-missing rate. 

Removing any component degrades performance, showing that the proposed designs provide complementary benefits. \textbf{Within SIC}, removing structural calibration causes a larger decrease than removing correspondence calibration, showing the importance of regularizing within-item cross-modal organization for reducing \textit{Cross-modal Structural Distortion}. Correspondence calibration provides gains by anchoring the completion space to cross-modal correspondence learned from genuinely observed modality pairs. 

\textbf{Within PRC}, removing the entire stage causes the largest degradation among the major module-level variants, confirming that structurally calibrated modality completion alone is insufficient for personalized ranking. Removing pseudo-missing alignment weakens representation-level adaptation of the recovered pathway, while removing the completion-aware item graph reduces connections available for preference propagation. These results support the representation- and relation-level design of PRC. The ablation results validate the sequential design of ~\shortname{}, where SIC calibrates the within-item cross-modal organization and PRC adapts representations and item neighborhoods to personalized ranking. 

\subsection{Performance on Incomplete Items}

\begin{figure}[t]
  \centering
  \includegraphics[width=0.95\columnwidth]{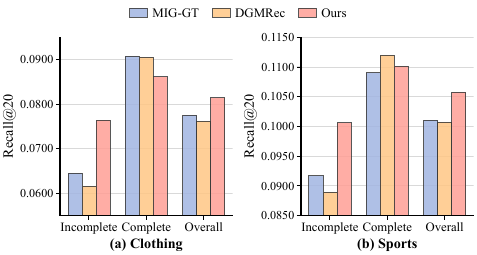}
  \caption{Performance across item groups.}
  \label{fig:complete_incomplete_items} 
  % \vspace{-0.2cm}
\end{figure}

\begin{figure}[t]
  \centering
  \includegraphics[width=0.95\columnwidth]{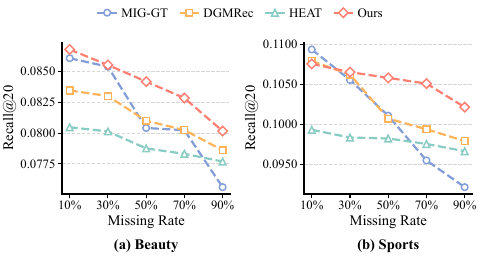}
  \caption{Performance under varying modality-missing rates.}
  \label{fig:missing_rate_trends}
\end{figure}

To directly evaluate the effectiveness of ~\shortname{} on modality-incomplete items, we divide test items into modality-incomplete and modality-complete groups according to the fixed item-level missing mask and evaluate them separately under the same ranking protocol. We compare ~\shortname{} with MIG-GT and DGMRec, the strongest general multimodal and missing-aware baselines, respectively. Figure~\ref{fig:complete_incomplete_items} reveals two main findings. 
(1) ~\shortname{} achieves substantially larger improvements on modality-incomplete items and consistently outperforms both baselines across datasets. On modality-complete items, the competing methods remain competitive, and ~\shortname{} is slightly weaker in some settings. Nevertheless, its gains on incomplete items outweigh these differences and yield the strongest overall performance, showing that the improvement of ~\shortname{} is primarily concentrated on items directly affected by modality missingness.
(2) The substantially larger gap over DGMRec on modality-incomplete items suggests that modality completion alone does not necessarily produce recommendation-effective representations. By calibrating the cross-modal organization of completed modalities and further adapting the recovered pathway and item neighborhoods to personalized ranking, ~\shortname{} provides more effective recommendation for incomplete items, supporting the need for recommendation-oriented adaptation after completion.

\subsection{Robustness under Varying Missing Rates}

To evaluate robustness under different degrees of modality incompleteness, we vary the proportion of modality-incomplete items from $10\%$ to $90\%$ and compare ~\shortname{} with MIG-GT, DGMRec, and HEAT. Figure~\ref{fig:missing_rate_trends} reports two main findings. (1) Completion-based methods degrade more slowly as the missing rate increases. MIG-GT performs strongly under mild missingness but declines rapidly as more items lose one modality, whereas DGMRec, HEAT, and ~\shortname{} show flatter trends. This suggests that estimating unavailable modality representations can partially compensate for lost content signals. (2) ~\shortname{} achieves the strongest performance among completion-based methods, especially under moderate-to-severe missingness. Its advantage grows with the missing rate, indicating lower sensitivity to severe modality incompleteness. This is consistent with the group-wise analysis, where its gains mainly arise from modality-incomplete items. The results support calibrating the cross-modal organization of completed modalities and adapting recovered representations and item relations to personalized ranking. Additional image-only and text-only missingness results in Appendix~\ref{app:single_modality_missingness} further demonstrate robustness across missing-modality types.

\subsection{Hyperparameter Sensitivity}

\begin{figure}[t]
  \centering
  \raisebox{-0.15cm}[\height][0pt]{%
    \includegraphics[width=0.9\columnwidth]{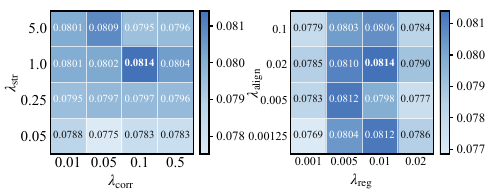}}
  \caption{Hyperparameter sensitivity of ~\shortname.}
  \label{fig:hyperparameter_sensitivity}
\end{figure}

Figure~\ref{fig:hyperparameter_sensitivity} examines the objective weights in SIC and PRC. For SIC, moderate values of $\lambda_{\mathrm{str}}$ and $\lambda_{\mathrm{corr}}$ generally yield stronger performance. A weak structural constraint provides limited cross-modal regularization, whereas an excessively large value may over-constrain the completion space. The effect of correspondence calibration also depends on structural calibration, suggesting that relation-pattern regularization and observed-pair supervision provide complementary signals.
A similar pattern appears for PRC. A small $\lambda_{\mathrm{align}}$ provides limited guidance for adapting the recovered pathway, while a large value may overemphasize matching the observed branch. The regularization weight must balance model capacity and overfitting. Overall, the results highlight the need to balance structural calibration and recommendation-oriented adaptation. Additional hyperparameter analyses are provided in Appendix~\ref{app:additional_hyperparameter_analysis}.

\section{Related Work}

\subsection{Multimodal Recommendation}

Multimodal recommendation incorporates item content such as images, text, and videos into collaborative preference modeling~\cite{he2016vbpr,wei2019mmgcn,ni2023content}. Early methods directly integrate modality features into recommendation objectives. VBPR~\cite{he2016vbpr} extends pairwise ranking with visual factors, while BM3~\cite{zhou2023bootstrap} and SLMRec~\cite{tao2023slmrec} introduce self-supervised learning across interaction and content views. Other methods construct modality-aware item graphs or propagate multimodal and collaborative signals, including LATTICE~\cite{zhang2021lattice}, MICRO~\cite{yi2022micro}, FREEDOM~\cite{zhou2023freedom}, PGL~\cite{yu2025pgl}, MMGCN~\cite{wei2019mmgcn}, GRCN~\cite{wei2020grcn}, MGCN~\cite{yu2023mgcn}, and LGMRec~\cite{guo2024lgmrec}. Recent studies further improve modality fusion and global dependency modeling~\cite{xu2024mentor,lin2024gume,lin2025cmdl,ong2025spectrum,hu2025miggt}.

These studies demonstrate the value of multimodal content beyond interaction-only recommenders such as LightGCN~\cite{he2020lightgcn} and SimGCL~\cite{yu2022simgcl}. However, most assume that all items have complete modality features \cite{wang2018lrmm,malitesta2024drop,kim2025disentangling}. Missing content weakens semantic evidence and disrupts modality-derived item relations, motivating incomplete multimodal recommendation.

\subsection{Incomplete Multimodal Recommendation}

Existing methods mainly address modality incompleteness through robust representation learning or explicit completion~\cite{bai2024multimodality,chen2025i3mrec,li2025generating}. Robust methods learn stable representations from available modalities. SiBraR~\cite{ganhor2024sibrar} maps different modality combinations into a shared space, MILK~\cite{bai2024multimodality} combines cross-modal alignment with invariant learning, and I$^{3}$-MRec~\cite{chen2025i3mrec} employs invariant risk minimization and an information bottleneck. However, their content evidence remains limited to observed modalities.

Completion-based methods estimate unavailable modality representations~\cite{ngiam2011multimodal,srivastava2012multimodal}. LRMM~\cite{wang2018lrmm} combines modality dropout with a multimodal autoencoder, while CI$^{2}$MG~\cite{lin2023contrastive} models intra-modality relations and inter-modality generation. Recent methods use conditional diffusion~\cite{li2025generating}, disentangled generation~\cite{kim2025disentangling}, graph-context retrieval~\cite{li2026robust}, or training-free feature propagation~\cite{malitesta2026trainingfree}. However, individual completion does not explicitly ensure coherent cross-modal organization or effectiveness for personalized ranking.

CalMRL~\cite{liu2026calibrated} infers shared semantics from observed modalities and uses imputation to recalibrate observed encoders against anchor shift in generic multimodal alignment. In contrast, ~\shortname{} preserves observed modalities as direct item content and uses their shared semantics to recover and calibrate the missing modality itself. SIC regularizes the recovered representation into a coherent within-item cross-modal structure, while PRC adapts only the recovery pathway to the ranking-supervised space and reconstructs completion-aware item relations for preference propagation.

\section{Conclusion}

In this work, we studied incomplete multimodal recommendation from the joint perspectives of modality completion and recommendation adaptation. We identified two successive challenges, \textit{Cross-modal Structural Distortion} and the \textit{Preference Adaptation Gap}. The former arises when individually recovered representations form inconsistent relations with other modalities of the same item, while the latter concerns whether completed representations and item relations are effective for personalized ranking. To address them, we proposed ~\fullname~(\shortname), a two-stage framework connecting modality completion with personalized ranking. SIC calibrates the within-item cross-modal organization of estimated missing-modality representations, while PRC adapts them in the recommendation space and supplements item relations for preference propagation. Extensive experiments on three real-world datasets under different modality-missing settings demonstrated the effectiveness and robustness of ~\shortname{}. Our evaluation is limited to Amazon benchmarks with visual and textual modalities and synthetically generated missingness. Extending ~\shortname{} to other domains, naturally occurring missing patterns, and settings with more than two modalities remains important future work. Overall, the results show that effective incomplete multimodal recommendation requires both structurally calibrated completion and recommendation-oriented adaptation.

\bibliographystyle{ACM-Reference-Format}
\newpage
\bibliography{references}
\newpage
\appendix

\section{Additional Method Details}
\label{app:method_details}

The shared-latent imputation backbone in SIC follows conjugate Gaussian latent-variable modeling~\cite{tipping1999probabilistic,bach2005probabilistic,klami2015group,ghojogh2021factor}.
This section provides the analytical posterior inference, pseudo-missing construction, graph propagation, and additional analysis omitted from the main text.

\subsection{SIC Posterior Inference}
\label{app:sic_inference}

Given the observed completion-space representations $\{\mathbf z_i^m\}_{m\in\mathcal O_i}$ and the generative parameters $\Theta_g=\{\mathbf W^m,\boldsymbol\mu^m,\sigma_m\}_{m\in\mathcal M}$, the posterior of the shared latent variable is:
\begin{equation}
q_i(\boldsymbol\beta_i) = p\left( \boldsymbol\beta_i \mid \{\mathbf z_i^m\}_{m\in\mathcal O_i}, \Theta_g \right) = \mathcal N(\mathbf b_i,\boldsymbol\Sigma_i),
\label{eq:app_sic_posterior}
\end{equation}
where
\begin{equation}
\boldsymbol\Sigma_i = \left[ \mathbf I_{d_\beta} + \sum_{m\in\mathcal O_i} \frac{(\mathbf W^m)^\top\mathbf W^m}{(\sigma_m)^2} \right]^{-1},
\label{eq:app_sic_covariance}
\end{equation}
and
\begin{equation}
\mathbf b_i = \boldsymbol\Sigma_i \sum_{m\in\mathcal O_i} \frac{(\mathbf W^m)^\top (\mathbf z_i^m-\boldsymbol\mu^m)}{(\sigma_m)^2}.
\label{eq:app_sic_mean}
\end{equation}
These expressions are obtained by collecting the quadratic terms in the log posterior and completing the square. Because the posterior depends only on the observed modality set $\mathcal O_i$, the same inference applies to different missing patterns.

\subsection{Pseudo-Missing Inference}
\label{app:pseudo_missing_inference}

For each eligible pair $(i,m)\in\mathcal E_{\mathrm{pm}}$, modality $m$ is temporarily removed from the observed set:
\begin{equation}
\mathcal O_i^{(-m)} = \mathcal O_i\setminus\{m\}.
\label{eq:app_pseudo_set}
\end{equation}
The corresponding posterior covariance and mean are:
\begin{equation}
\boldsymbol\Sigma_i^{(-m)} = \left[ \mathbf I_{d_\beta} + \sum_{n\in\mathcal O_i^{(-m)}} \frac{(\mathbf W^n)^\top\mathbf W^n}{(\sigma_n)^2} \right]^{-1},
\label{eq:app_pseudo_covariance}
\end{equation}
and
\begin{equation}
\mathbf b_i^{(-m)} = \boldsymbol\Sigma_i^{(-m)} \sum_{n\in\mathcal O_i^{(-m)}} \frac{(\mathbf W^n)^\top (\mathbf z_i^n-\boldsymbol\mu^n)}{(\sigma_n)^2}.
\label{eq:app_pseudo_mean}
\end{equation}
The pseudo-recovered representation is:
\begin{equation}
\widehat{\mathbf z}_i^{m,(-m)} = \frac{\mathbf W^m\mathbf b_i^{(-m)}+\boldsymbol\mu^m}{\left\| \mathbf W^m\mathbf b_i^{(-m)}+\boldsymbol\mu^m \right\|_2+\epsilon}.
\label{eq:app_pseudo_recovery}
\end{equation}
Pseudo-recovered and genuinely missing representations therefore follow the same posterior inference and target-modality generation process.

\subsection{Completion-Aware Graph Propagation}
\label{app:graph_propagation}

For modality $m$, collaborative and completed content similarities are first fused as:
\begin{equation}
\mathbf S^{m,\mathrm{fuse}} = (1-\eta)\mathbf S^{\mathrm{cf}} + \eta\mathbf S^m.
\label{eq:app_fused_similarity}
\end{equation}
We then retain the row-wise Top-$k$ entries and symmetrize the resulting graph:
\begin{equation}
\overline{\mathbf G}_{\mathrm{item}}^m = \mathcal T_k \left( \mathbf S^{m,\mathrm{fuse}} \right), \qquad \mathbf G_{\mathrm{item}}^m = \mathcal S \left( \overline{\mathbf G}_{\mathrm{item}}^m \right),
\label{eq:app_item_graph}
\end{equation}
where $\mathcal T_k(\cdot)$ denotes row-wise Top-$k$ sparsification and $\mathcal S(\cdot)$ denotes graph symmetrization.

The modality-specific augmented adjacency matrix is:
\begin{equation}
\mathbf A^m = \begin{bmatrix} \mathbf 0 & \mathbf R\\ \mathbf R^\top & \mathbf G_{\mathrm{item}}^m \end{bmatrix}.
\label{eq:app_augmented_graph}
\end{equation}
Following LightGCN~\cite{he2020lightgcn}, we apply symmetric normalization:
\begin{equation}
\widehat{\mathbf A}^m = (\mathbf D^m)^{-\frac{1}{2}} \mathbf A^m (\mathbf D^m)^{-\frac{1}{2}}, \qquad \mathbf D^m = \operatorname{Diag}(\mathbf A^m\mathbf 1).
\label{eq:app_normalized_graph}
\end{equation}

The initial node representations are:
\begin{equation}
\mathbf H^{m,(0)} = \begin{bmatrix} \mathbf E_U^{(0)}\\ \mathbf E_I^{m,(0)} \end{bmatrix}, \qquad \mathbf E_I^{m,(0)} = [\mathbf s_1^m,\ldots,\mathbf s_{N_i}^m]^\top.
\label{eq:app_initial_nodes}
\end{equation}
Layer-wise propagation is performed as:
\begin{equation}
\mathbf H^{m,(l+1)} = \widehat{\mathbf A}^m \mathbf H^{m,(l)}.
\label{eq:app_propagation}
\end{equation}
The representations from all layers are uniformly aggregated:
\begin{equation}
\mathbf H^m = \frac{1}{L+1} \sum_{l=0}^{L} \mathbf H^{m,(l)}.
\label{eq:app_layer_aggregation}
\end{equation}

Let $\mathbf p_u^m$ and $\mathbf q_i^m$ denote the propagated user and item representations under modality $m$.
The final representations are:
\begin{equation}
\mathbf h_u = \frac{1}{N_m} \sum_{m\in\mathcal M} \mathbf p_u^m, \qquad \mathbf h_i = \frac{1}{N_m} \sum_{m\in\mathcal M} \mathbf q_i^m.
\label{eq:app_modality_fusion}
\end{equation}
During PRC training, the SIC parameters, completed modality representations, and completion-aware item graphs remain fixed.

\subsection{Bimodal Interpretation of Structural Calibration}
\label{app:bimodal_structure}

For two normalized modality representations, let $c_i=(\tilde{\mathbf z}_i^1)^\top\tilde{\mathbf z}_i^2$.
Their Gram matrix is:
\begin{equation}
\mathbf G_i = \begin{bmatrix} 1 & c_i\\ c_i & 1 \end{bmatrix},
\end{equation}
with eigenvalues $\lambda_{i,1}=1+|c_i|$ and $\lambda_{i,2}=1-|c_i|$.
Consequently, the leading-eigenvalue component of $\mathcal L_{\mathrm{str}}$ increases with $|c_i|$ in the bimodal setting.

This component should not be interpreted as directly matching the ground-truth cross-modal geometry or recovering a rich item-specific structural configuration.
With two modalities, the within-item structure contains only one pairwise degree of freedom.
The leading-eigenvalue component therefore acts as a relation-concentration regularizer that discourages near-orthogonal configurations.
It does not independently determine the sign or semantic correctness of the pairwise relation.

In the complete SIC objective, $\mathcal L_{\mathrm{rep}}$ constrains the conditional estimate using the observed representations, while $\mathcal L_{\mathrm{corr}}$ provides same-item correspondence supervision from genuinely observed modality pairs.
The structural component is therefore optimized jointly with the imputation and correspondence objectives rather than used as an isolated recovery criterion.

When $N_m=2$, the principal eigenvector is constant within each fixed similarity-sign region.
The principal-direction component therefore provides no additional continuous gradient in our bimodal experiments, and the leading-eigenvalue component supplies the effective structural signal.
The principal-direction formulation is retained for settings with more than two modalities, where nontrivial modality-wise directional configurations can arise.
Its empirical benefit in such settings remains to be investigated.

\subsection{Model Analysis}
\label{app:model_analysis}

\subsubsection{Computational complexity.}
Let $|\mathcal E_R|=\|\mathbf R\|_0$, and let $Q$ and $L$ denote the numbers of distinct nonempty observed-modality patterns and propagation layers, respectively. The main costs of SIC are:
\begin{equation}
\begin{aligned}
\text{Projection:}\quad
&\mathcal O\left(
\sum_{i\in\mathcal I}
\sum_{m\in\mathcal O_i}
d_md_c
\right),\\
\text{Posterior inference:}\quad
&\mathcal O\left(
N_md_cd_\beta^2
+Qd_\beta^3
+\sum_{i\in\mathcal I}|\mathcal O_i|d_cd_\beta
+N_id_\beta^2
\right),\\
\text{Structural calibration:}\quad
&\mathcal O\left(
N_iN_m^2d_c
+N_iN_m^3
+N_i|\mathcal B|N_m
\right).
\end{aligned}
\label{eq:app_sic_complexity}
\end{equation}
Here, $(\mathbf W^m)^\top\mathbf W^m$ is reused by items sharing the same observed-modality pattern, and the final structural-calibration term arises from batch-wise principal-direction comparison. Correspondence calibration is linear in the number of sampled observed cross-modal pairs.

Before PRC training, exact content-based neighbor construction and collaborative co-occurrence computation require:
\begin{equation}
\mathcal O\left(
N_mN_i^2d_c
+\sum_{u\in\mathcal U}\delta_u^2
\right),
\qquad
\delta_u=\sum_iR_{ui}.
\label{eq:app_graph_construction_complexity}
\end{equation}
In our implementation, $\mathbf R$ is stored sparsely, collaborative co-occurrence is computed by sparse matrix multiplication, and content similarities are evaluated in blocks. Only the Top-$k$ neighbors are retained and the resulting graphs are symmetrized, so the full dense $N_i\times N_i$ similarity matrix is not stored simultaneously. Each modality-specific graph contains $\mathcal O(N_ik)$ retained edges and is constructed once for reuse throughout PRC training.

After graph construction, the per-epoch PRC costs are:
\begin{equation}
\begin{aligned}
\text{Graph propagation:}\quad
&\mathcal O\left(
LN_m(|\mathcal E_R|+kN_i)d
\right),\\
\text{Ranking and alignment:}\quad
&\mathcal O\left(
\bigl(
|\mathcal D_{\mathrm{BPR}}|
+|\mathcal E_{\mathrm{pm}}|
\bigr)d
\right).
\end{aligned}
\label{eq:app_prc_complexity}
\end{equation}
Thus, the quadratic term is incurred only during exact content-neighbor construction, whereas repeated propagation is linear in the numbers of interaction and retained item-graph edges. Exact construction may still be expensive for million-scale item collections, for which approximate nearest-neighbor or distributed graph construction can be adopted.

\subsubsection{Empirical efficiency.}

Table~\ref{tab:training_efficiency} reports per-epoch training time and full-ranking recommendation inference time under the same experimental setting.
Because \shortname{} is optimized sequentially, SIC and PRC are reported separately.
SIC requires $5.9$ seconds per training epoch.
After the completed representations and item graphs are constructed, PRC requires $12.5$ seconds per epoch and $5.6$ seconds for full-ranking inference.
The one-time missing-modality completion pass takes $3.9\times10^{-3}$ seconds.

PRC has a slightly higher per-epoch training cost than DGMRec but is faster than HEAT and MoDiCF.
Its full-ranking inference is faster than DGMRec and MoDiCF, while HEAT remains the fastest.
These results show that recommendation-space adaptation and sparse graph propagation introduce moderate per-epoch overhead.
The one-time exact item-graph construction cost is not included in the reported PRC per-epoch time and is discussed separately in the complexity analysis.

\begin{table}[t]
    \centering
    \caption{Per-epoch training and inference/preprocessing time under the same experimental setting. The SIC entry denotes the one-time missing-modality completion pass, whereas the PRC entry denotes full-ranking recommendation inference.}
    \label{tab:training_efficiency}
    \small
    \setlength{\tabcolsep}{3pt}
    \begin{tabular}{@{}lcc@{}}
        \toprule
        Method or Stage & Train/Epoch (s) & \shortstack{Inference /\\Preprocessing (s)} \\
        \midrule
        DGMRec & 10.1 & 10.8 \\
        HEAT & 20.2 & 4.4 \\
        MoDiCF & 761.0 & 33.2 \\
        \shortname{}: SIC & 5.9 & $3.9\times10^{-3}$ \\
        \shortname{}: PRC & 12.5 & 5.6 \\
        \bottomrule
    \end{tabular}
\end{table}

\subsection{Relation to CalMRL and Task-Specific Distinction}
\label{app:relation_to_calmrl}

Although ~\shortname{} builds on the calibrated representation recovery principle of CalMRL~\cite{liu2026calibrated}, the two methods differ fundamentally in the optimization role of the recovered modality. CalMRL studies missing modalities in generic multimodal representation learning, where anchor shift denotes the deviation between the alignment direction induced by the complete modality set and that induced by the observed subset. It estimates the missing representation to compensate for its absent contribution to the virtual alignment anchor and thereby recalibrate the observed modality encoders. Recovery thus primarily serves as an auxiliary mechanism for improving the available modalities.

~\shortname{} shifts the role of recovery. It operates on pretrained item features and preserves genuinely observed modalities as direct item content. Their shared semantics are instead used to estimate the unavailable modality, whose recovered representation is retained as explicit content and must itself become structurally reliable. Accordingly, \textit{Cross-modal Structural Distortion} is related to but distinct from anchor shift: it concerns whether the recovered modality forms a coherent within-item organization with the observed modalities, rather than whether it restores an alignment direction for optimizing observed encoders. SIC therefore specializes calibrated recovery toward improving the recovered modality and organizing the completed item content.

This shift further exposes the \textit{Preference Adaptation Gap}: structural reliability alone does not ensure that recovered content is compatible with the ranking-supervised representation space, while modality missingness also disrupts content-derived item neighborhoods. PRC addresses this gap by aligning the recovery pathway with its stop-gradient observed counterpart through same-item, same-modality pseudo-missing supervision, and by combining completion-aware content relations with collaborative evidence for preference propagation. Removing PRC produces the largest module-level degradation, while removing either pseudo-missing alignment or the completion-aware item graph also consistently reduces performance. These results show that ~\shortname{}'s contribution lies not in latent-variable imputation alone, but in redirecting calibrated recovery toward improving the missing modality itself and enabling its effective participation in personalized recommendation.

\subsection{Model Limitations}

Our evaluation is limited to three Amazon datasets containing visual and textual modalities.
Although these datasets are widely used in multimodal recommendation, they belong to the same platform and domain family.
The current results therefore do not establish generalization to other domains or to settings involving audio, video, attributes, or more than two modalities.

This limitation is particularly relevant to the principal-direction component of structural calibration.
In the bimodal setting, the within-item structure contains only one pairwise relation, and the principal-direction component provides no additional continuous gradient within a fixed similarity-sign region.
Its role in settings with richer modality configurations remains to be empirically validated.

The main experiments also use synthetically generated modality-missing masks.
Real-world missingness may depend on item popularity, category, content quality, or interaction sparsity.
Evaluating \shortname{} under naturally occurring and non-random missing mechanisms is an important direction for future work.

Finally, correspondence calibration and pseudo-missing adaptation require training items with at least two genuinely observed modalities.
When eligible paired observations are extremely scarce or unavailable, these objectives provide limited or no supervision.
The effectiveness of \shortname{} under such settings remains to be investigated.

\section{Experimental Details}

\subsection{Baselines and Evaluation}
\label{app:baselines}

We compare \shortname{} with eleven representative baselines covering collaborative filtering, general multimodal recommendation, and missing-aware multimodal recommendation.
\begin{itemize}[leftmargin=*,topsep=2pt,itemsep=1pt,parsep=0pt]
  \item \textbf{LightGCN}~\citep{he2020lightgcn} simplifies graph convolution to neighborhood aggregation and learns collaborative representations from user--item interactions.
  \item \textbf{SimGCL}~\citep{yu2022simgcl} injects noise into node embeddings to construct contrastive views without explicit graph augmentation.
  \item \textbf{VBPR}~\citep{he2016vbpr} extends pairwise ranking by incorporating visual item features.
  \item \textbf{BM3}~\citep{zhou2023bootstrap} learns multimodal representations through bootstrap self-supervision over interactions and item content.
  \item \textbf{PGL}~\citep{yu2025pgl} uses principal subgraphs to model global co-occurrence and local personalized information.
  \item \textbf{MIG-GT}~\citep{hu2025miggt} combines modality-independent GNN receptive fields with a sampling-based global Transformer.
  \item \textbf{I$^3$-MRec}~\citep{chen2025i3mrec} addresses modality incompleteness through invariant representation learning and an information bottleneck.
  \item \textbf{MILK}~\citep{bai2024multimodality} learns modality-invariant preferences across heterogeneous missing environments.
  \item \textbf{MoDiCF}~\citep{li2025generating} combines diffusion-based modality completion with counterfactual recommendation.
  \item \textbf{DGMRec}~\citep{kim2025disentangling} disentangles modality-shared and modality-specific information before generating missing features.
  \item \textbf{HEAT}~\citep{malitesta2026trainingfree} performs training-free feature propagation over an item--item graph to impute unavailable modalities.
\end{itemize}

We use official implementations when available and reproduce the remaining baselines following their original papers.
Hyperparameters are selected on the validation set, and all methods use the same data splits and modality-missing masks.

Following the full-ranking protocol commonly used in graph recommendation~\cite{he2020lightgcn,yu2022simgcl}, we report Recall@$K$ and NDCG@$K$ for $K\in\{10,20\}$.
Recall@$K$ measures the proportion of held-out interactions retrieved in the top-$K$ list, while NDCG@$K$ assigns greater weight to items ranked at higher positions~\cite{jarvelin2002ndcg}.
For each user, all items except those observed during training are used as ranking candidates, and the results are averaged over test users.

Each method is run five times with different random seeds while keeping the data split and modality-missing mask fixed.
Within each run, all methods use the same experimental setting.
We report the mean performance over the five runs.
Because the data split and missing mask remain fixed, differences across runs arise from random initialization and stochastic optimization rather than different missing masks.
Statistical significance between ~\shortname{} and the strongest baseline is evaluated using a paired two-sided $t$-test.

\subsection{Hyperparameter Settings}
\label{app:hyperparameter_settings}

The completion and recommendation dimensions are both set to 64, and the mini-batch size is fixed at 2,048.
We tune the learning rate over $\{10^{-4},5\times10^{-4},10^{-3}\}$ and the $L_2$ regularization coefficient over $\{10^{-5},10^{-4},10^{-3}\}$.
The structural-calibration temperature $\tau$ is selected from $\{0.05,0.1,0.2,0.5\}$.

The SIC weights $\lambda_{\mathrm{str}}$ and $\lambda_{\mathrm{corr}}$ are selected according to validation performance from the candidate values examined in the sensitivity analysis.
For PRC, we tune $\lambda_{\mathrm{align}}$ over $\{10^{-4},1.25\times10^{-3},5\times10^{-3},2\times10^{-2},10^{-1}\}$ and $\lambda_{\mathrm{reg}}$ over $\{10^{-3},5\times10^{-3},10^{-2},2\times10^{-2}\}$.

The item-graph hyperparameters are selected from:
\begin{equation}
k\in\{1,3,5,10,20\}, \qquad \eta\in\{0.1,0.25,0.4,0.7,0.9\}.
\end{equation}
All hyperparameters are selected according to validation Recall@20 under the same data split and modality-missing setting used for final evaluation.
We apply early stopping when validation Recall@20 does not improve for 30 consecutive epochs and use the best validation checkpoint for testing.

\begin{figure*}[t]
  \centering
  \subfloat[Clothing]{
    \includegraphics[width=0.3\textwidth]{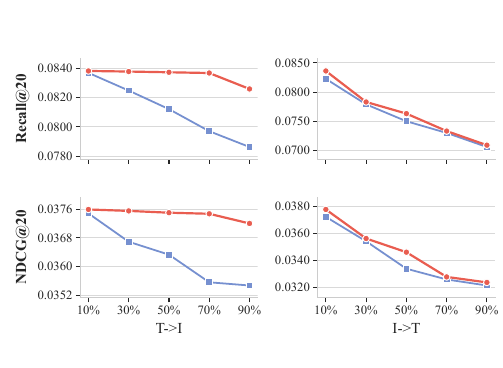}
  }\hfill
  \subfloat[Sports]{
    \includegraphics[width=0.3\textwidth]{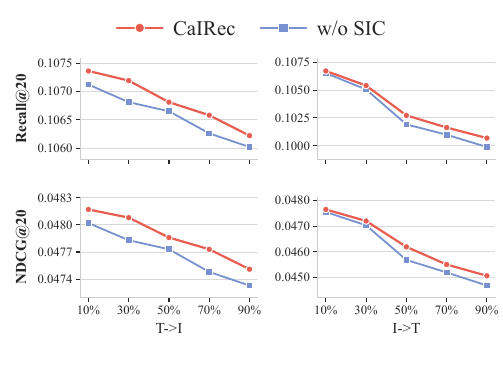}
  }\hfill
  \subfloat[Beauty]{
    \includegraphics[width=0.3\textwidth]{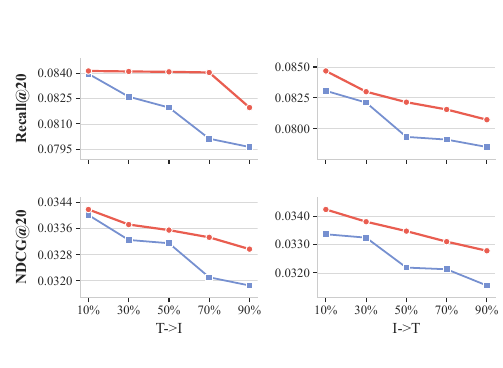}
  }
  \caption{Performance of \shortname{} and its w/o SIC variant under varying single-modality missing rates. $T{\to}I$ denotes image completion from text, while $I{\to}T$ denotes text completion from images.}
  \label{fig:single_modality_missing_rates}
\end{figure*}

\section{Further Analysis}
\label{app:additional_experiments}

\subsection{Performance under Single-Modality Missingness}
\label{app:single_modality_missingness}

The main experiments randomly select whether image or text is unavailable for each modality-incomplete item.
We additionally consider two controlled missing directions.
$T{\to}I$ removes images and estimates them from text, while $I{\to}T$ removes text and estimates it from images.
For each direction, we vary the missing rate from $10\%$ to $90\%$ and compare ~\shortname{} with its w/o SIC variant.

Figure~\ref{fig:single_modality_missing_rates} shows that SIC consistently improves Recall@20 and NDCG@20 under both missing directions.
The advantage generally increases at higher missing rates, particularly under $T{\to}I$ on Clothing and Beauty.
This result shows that SIC remains beneficial when the missing modality type is fixed rather than randomly selected.

Performance is generally lower under $I{\to}T$ than under $T{\to}I$, especially at high missing rates.
This difference may reflect the greater recommendation utility of textual content, the difficulty of estimating it from images, or both.
The current experiment does not isolate these explanations.
Overall, the results show that the effect of modality incompleteness depends on which modality is unavailable, while SIC provides consistent gains in both directions.

\subsection{Additional Results under Varying Missing Rates}
\label{app:additional_missing_rate_results}

\begin{figure*}[t]
  \centering
  \includegraphics[width=0.90\textwidth]{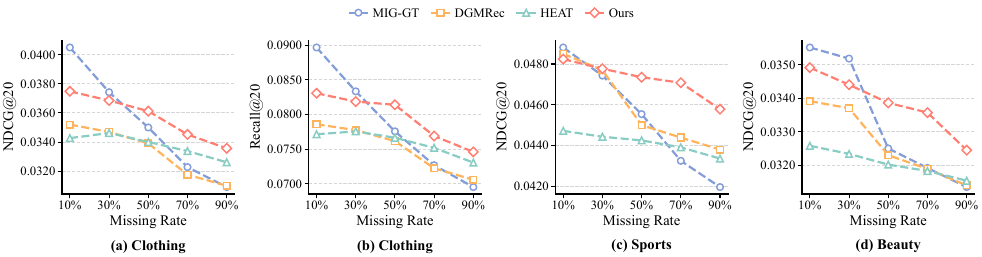}
  \caption{Additional recommendation results under varying modality-missing rates.}
  \label{fig:missing_rate_trends_supplementary}
\end{figure*}

Figure~\ref{fig:missing_rate_trends_supplementary} reports additional results under varying modality-missing rates.
Completion-based methods generally degrade more slowly than MIG-GT as the missing rate increases, indicating that explicit modality estimation can partially compensate for the loss of original content.
Among these methods, \shortname{} maintains the strongest or near-strongest performance and shows clearer advantages under severe missingness.
These observations are consistent with the robustness analysis in the main text.

\subsection{Stage-Wise Optimization}
\label{app:staged_training}

SIC and PRC optimize different representation spaces.
SIC learns conditional modality imputation and within-item cross-modal organization, whereas PRC learns recommendation-space transformations and preference propagation.
Under joint optimization, ranking gradients can update the SIC parameters through the recovered branch, causing the completion process and recommendation model to change simultaneously.
Stage-wise optimization instead fixes the completion process before learning recommendation-oriented adaptation.

\begin{table}[t]
  \centering
  \caption{Effect of stage-wise optimization on Clothing. $\Delta$ reports the relative change from the stage-wise strategy.}
  \label{tab:staged_training}
  \footnotesize
  \setlength{\tabcolsep}{3.2pt}
  \renewcommand{\arraystretch}{1.08}
  \begin{tabular}{lccc}
    \toprule
    Training Strategy & Recall@20 & NDCG@20 & $\Delta$ Recall / NDCG \\
    \midrule
    Stage-wise (Ours)             & \textbf{0.0814} & \textbf{0.0361} & -- \\
    Warm-started Joint            & 0.0753 & 0.0337 & $-7.49\%$ / $-6.65\%$ \\
    From-scratch Joint (Rec-only) & 0.0753 & 0.0336 & $-7.49\%$ / $-6.93\%$ \\
    From-scratch Joint (Rec+SIC)  & 0.0756 & 0.0339 & $-7.13\%$ / $-6.09\%$ \\
    \bottomrule
  \end{tabular}
\end{table}

Table~\ref{tab:staged_training} shows that stage-wise optimization outperforms all three joint-training variants on Clothing.
Warm-started joint training remains weaker despite using the same SIC initialization, suggesting that pretraining alone does not explain the improvement.
The two from-scratch variants perform similarly, while retaining the SIC objectives yields only a modest gain over recommendation-only joint training.

These results are consistent with the rationale that fixing SIC provides a more stable input space for PRC and reduces coupling between completion and ranking objectives.
However, this comparison does not establish gradient interference as the only cause of the performance difference.

\subsection{Plug-In Evaluation of SIC}
\label{sec:plugin_evaluation}

\begin{figure}[t]
  \centering
  \includegraphics[width=\columnwidth]{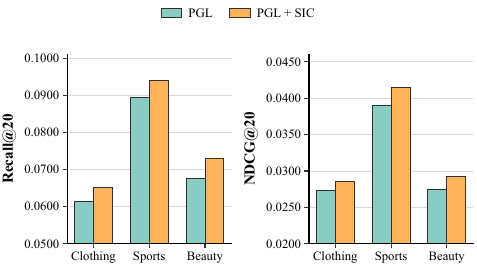}
  \caption{Performance of PGL and PGL+SIC on the three Amazon datasets under a $50\%$ modality-missing rate.}
  \label{fig:plugin_pgl}
\end{figure}

We examine whether SIC can be incorporated into an existing multimodal recommender without modifying its recommendation backbone.
We apply SIC to PGL and denote the resulting variant as PGL+SIC.

As shown in Figure~\ref{fig:plugin_pgl}, PGL+SIC consistently outperforms PGL on all three datasets in Recall@20 and NDCG@20.
The gains suggest that the completed representations produced by SIC can also support a recommendation backbone not specifically designed for modality missingness.
This result provides additional evidence that SIC can serve as a plug-in completion component.

\subsection{Additional Hyperparameter Experiments}
\label{app:additional_hyperparameter_analysis}

\begin{figure}[t]
  \centering
  \includegraphics[width=\columnwidth]{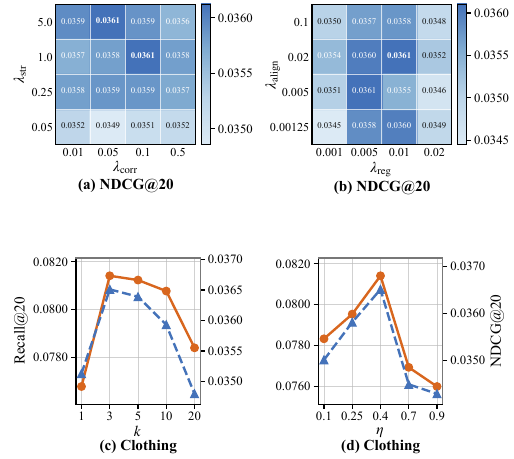}
  \caption{Additional hyperparameter sensitivity on Clothing.}
  \label{fig:sensitivity_hyperparameters}
  \Description{Two NDCG@20 heatmaps for the SIC and PRC objective weights and two line charts for the graph parameters $k$ and $\eta$ on Clothing.}
\end{figure}

Figure~\ref{fig:sensitivity_hyperparameters} provides complementary NDCG@20 results for the SIC and PRC objective weights and examines the graph parameters $k$ and $\eta$ on Clothing.
Here, $k$ controls the number of neighbors retained in each modality-specific item graph, while $\eta$ balances collaborative and completed content similarities.
We vary one graph hyperparameter at a time while fixing the remaining settings.

Figures~\ref{fig:sensitivity_hyperparameters}(a)--(b) show broad high-performing regions under moderate SIC and PRC objective weights.
Figures~\ref{fig:sensitivity_hyperparameters}(c)--(d) exhibit increase-then-decrease trends for the graph parameters.
A very small $k$ provides limited item neighborhoods, whereas a large value introduces less relevant neighbors.
Similarly, a moderate $\eta$ balances collaborative and completed content relations.
These results show that \shortname{} is stable around suitable settings but can be affected by excessively weak or strong calibration and graph signals.

\begin{figure}[t]
  \centering
  \includegraphics[width=\columnwidth]{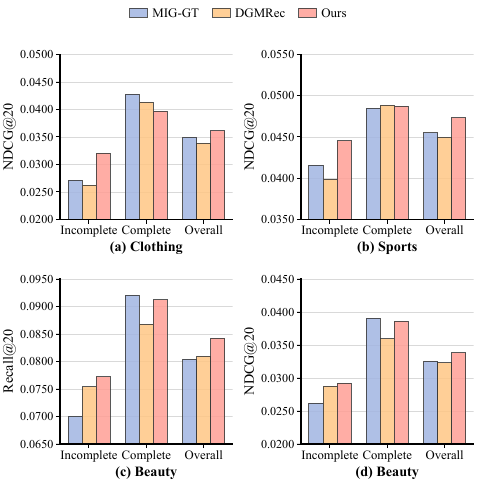}
  \caption{Additional performance comparison on modality-incomplete, modality-complete, and overall test-item groups.}
  \label{fig:complete_incomplete_items_supplementary}
\end{figure}

\subsection{Additional Results on Modality-Incomplete Items}
\label{app:supplementary_group_results}

Figure~\ref{fig:complete_incomplete_items_supplementary} reports additional group-wise results under the same setting as the main text.
Consistent with Figure~\ref{fig:complete_incomplete_items}, \shortname{} achieves its clearest gains on modality-incomplete items, while competing methods remain competitive on modality-complete items.
This shows that the overall improvement is primarily associated with items directly affected by modality missingness.
The advantage over DGMRec on incomplete items further suggests that explicit modality completion alone may not be sufficient to produce recommendation-effective representations.

\end{document}